\documentclass[journal]{IEEEtran}

\ifCLASSINFOpdf
\else
   \usepackage[dvips]{graphicx}
\fi
\usepackage{url}

\newcommand{\vx}{\mathbf{x}}
\newcommand{\vy}{\mathbf{y}}
\usepackage{color}
\usepackage{amsmath}
\usepackage{subcaption} 
\usepackage{balance}

\usepackage{amssymb}

\usepackage{graphicx}
\usepackage{booktabs,multirow}

\begin{document}
\title{Discovering shared interpretable operations in image compression autoencoders}
\author{Caroline Mazini Rodrigues, Nicolas Keriven, and Thomas Maugey \\Univ. Rennes, Inria, CNRS, IRISA, Rennes, France
\thanks{The authors acknowledge fundings of France 2030, PEPR IA, ANR-23-PEIA-0008 and European Union ERC-2024-STG-101163069 MALAGA.}}

\maketitle

\begin{abstract}
With the increasing adoption of deep learning for applications such as image compression, improvements in the rate-distortion trade-off have been achieved at the cost of increasingly larger and more opaque ``black-box'' models. Autoencoders are among the most widely used architectures for this task; however, without a clear understanding of their internal behavior, these models tend to grow in complexity to achieve more performance gains. In this paper, we investigate whether universal behaviors can be detected from the internal operations of bias-free autoencoders through Jacobian analysis. If such behaviors exist, they may be extracted to design low-complexity image compression models inspired by high-complexity deep learning architectures.
\end{abstract}

\begin{IEEEkeywords}
Interpretability, Image compression, Explainable Artificial Intelligence, Jacobian analysis, Frugality
\end{IEEEkeywords}

\IEEEpeerreviewmaketitle

\section{Introduction}

Autoencoders are widely used for image compression tasks, whose main objective is to reduce the amount of transmitted information (rate) while preserving a low reconstruction error (distortion) at the receiver side~\cite{balle:iclr:2017:factorized, balle:iclr:2018:hyper}. These models have progressively replaced handcrafted transformations, such as the Discrete Cosine Transform (DCT), because they can achieve better rate–distortion trade-offs~\cite{yang:2023:trends:intro1}. However, improvements in compression performance are often obtained by increasing the size and complexity of the models. This can limit their applicability in real-world scenarios with constrained hardware and may also lead to higher energy consumption~\cite{Gilbert:2024:ahn, yang:iccv:2023:shallow:decoder,liang:2020:itj}.

Currently, larger deep learning models achieve high performance by adapting to their input data. However, the knowledge learned during training remains hidden within these models, which are often treated as ``black boxes'' due to their complex nonlinear operations and large scale~\cite{hassija:2023:interpreting}. If this hidden knowledge could be better understood, we may use it more directly, reducing the need for some of the original model operations and potentially improving efficiency.

Some studies, such as Mohan~\textit{et~al.}~\cite{mohan:2020:iclr} and Kadkhodaie~\textit{et~al.}~\cite{kadkhodaie:2024:iclr}, propose interpreting autoencoder models as a set of input-adapted linear operations, similar to handcrafted linear filters.
However, these explanations are mainly local, since they focus on the model's adaptation to individual images~\cite{lundberg:2017:shap, ribeiro:2016:lime, RODRIGUES:2024:prl}. This contrasts with global explanations, which aim to identify universal mechanisms consistently used by the model across different input data~\cite{ghorbani2019ace, RODRIGUES:2024:infoSciences,rodrigues2026xaitrees}.

\begin{figure}[!ht]
\center{\includegraphics[width=0.9\linewidth]{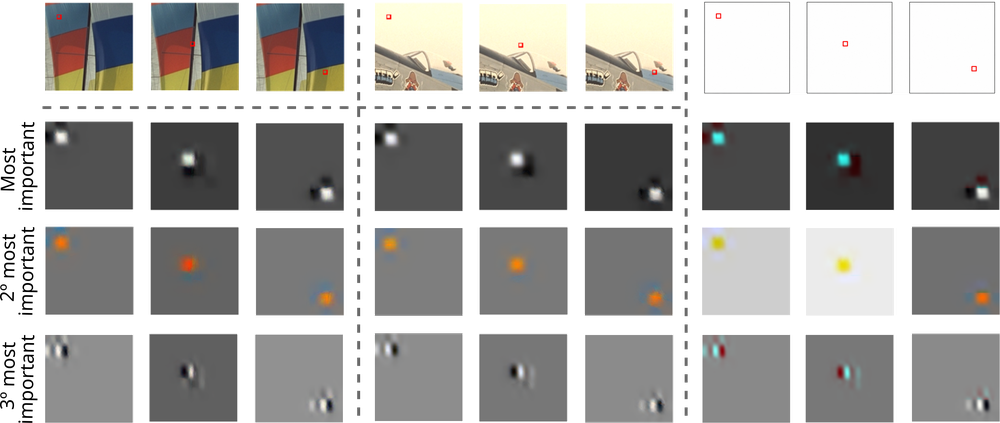}
}
\caption{\small{\textbf{Filtering operations by pixel in a compression autoencoder.} In the first row, we show pixel positions in two images filtered by the model. We present a schematic of the filters' construction in Figure~\ref{fig:diagram}. The same three positions are used for all images. 
 We show the three most important filters for preserving reconstruction quality, as identified through Jacobian analysis, and observe consistent behavior across both images for the most influential features. The white image represents our proposed aggregation of Jacobian-derived filters across multiple images, illustrating that a similar filtering signal is preserved. The color variations have little impact on quality, as shown in Section~\ref{sec:comp_3ch} and in the channel ablation in Appendix~J.
}}
\label{fig:individual_comb_filters}
\end{figure}

Inspired by the idea of describing a model as a set of input-adapted linear filters, we investigate the universality of these filters across different images in the compression task. \textcolor{black}{While previous studies have explored Jacobian-based interpretations of neural networks, our focus is on understanding how these filters vary across images and identifying common operations.} Figure~\ref{fig:individual_comb_filters} shows the similarities of image-adapted filters extracted from a compression model for two different images and raises the questions: \textit{does the model shows a general behavior across images, and at what point does it become image-specific?} Our intuition is that, if some filters are common across images, they could be used to develop low-complexity, deep-learning-inspired compression tools.
\vspace{-3mm}

\vspace{-5mm}

\section{Do compression models learn universality?}
In this work, \textit{we investigate whether deep learning autoencoders trained for image compression learn universal compression behaviors that generalize across images, or whether the model must fully adapt to each specific input.}

Our approach is motivated by Mohan~\textit{et~al.}~\cite{mohan:2020:iclr}, who use Jacobian matrices derived from bias-free autoencoders with ReLU activations to explain the model's behavior as a set of image-adaptable linear filters. Bias-free neural networks can be expressed as follows $f_{\mathrm{BF}}(\vx) = W_L \, \sigma\bigl(W_{L-1} \cdots \sigma(W_1 \vx)\bigr) = \mathcal{J}_{\vx} \vx$, where $\mathcal{J}_{\vx}$ denotes the Jacobian matrix evaluated at the input $\vx \in \mathcal{D}$, with $\mathcal{D}$ representing the dataset. The matrices $W_i$, for $i \in \{1, \dots, L\}$, correspond to the weight parameters of the $i$-th layer and $\sigma$ are ReLU activations. One advantage of using these bias-free networks with ReLU is that the model's exact filtering operations can be interpreted via per-image Jacobians. Specifically, by examining the rows of the Jacobian, we can extract linear, adaptive filters specific to each input.
Motivated by this Jacobian interpretation, we extend the analysis to compression models. Conventional autoencoders for compression~\cite{Jamil:2023:EAAI} optimize the rate-distortion trade-off:
\begin{equation}
\min_{g_{\mathrm{enc}}, g_{\mathrm{dec}}}
\mathbb{E}_\vx \left[ -\log p(g_{\mathrm{enc}}(\vx))
+
\lambda
\left\| \vx - g_{\mathrm{dec}}(g_{\mathrm{enc}}(\vx)) \right\|_2^2 \right],
\label{eq:rd_loss}
\end{equation}
with $g_{\mathrm{enc}}(\cdot)$ the encoder and $g_{\mathrm{dec}}(\cdot)$ the decoder. 
These two modules play roles analogous to a DCT matrix and its inverse, respectively, but are implemented using learned non-linear operations.
While linear transforms such as the DCT can be interpreted as signal decompositions, neural networks like autoencoders for image compression are black boxes.

In compression autoencoders, $g_{\mathrm{enc}}(\cdot)$ operates on the sender side and $g_{\mathrm{dec}}(\cdot)$ on the receiver side. For this reason, when explaining these models using the Jacobians, we must analyze the two components, $g_{\mathrm{enc}}(\cdot)$ and $g_{\mathrm{dec}}(\cdot)$, separately. Therefore, we can write the two modules as  in Equations~\eqref{eq:bf_encoder} and \eqref{eq:bf_decoder}.
\begin{equation}
    g_{\mathrm{enc}}(\vx) = W_M^{\mathrm{enc}} \, \sigma\Bigl(W_{M-1}^{\mathrm{enc}} \cdots \sigma(W_1^{\mathrm{enc}} \vx)\Bigr) = \mathcal{J}_{\mathrm{enc},\vx} \, \vx,
    \label{eq:bf_encoder}
\end{equation}
\begin{equation}
    g_{\mathrm{dec}}(\vy) = W_N^{\mathrm{dec}} \, \sigma\Bigl(W_{N-1}^{\mathrm{dec}} \cdots \sigma(W_1^{\mathrm{dec}} \vy)\Bigr) = \mathcal{J}_{\mathrm{dec},\vy} \, \vy,
    \label{eq:bf_decoder}
\end{equation}
where $M$ and $N$ denote the number of layers in the encoder and decoder, respectively, and the transmitted latent representation of the input image $\vx$ is given by $\vy = g_{\mathrm{enc}}(\vx)$. 

\begin{figure}[!ht]
\centerline{\includegraphics[width=0.8\linewidth]{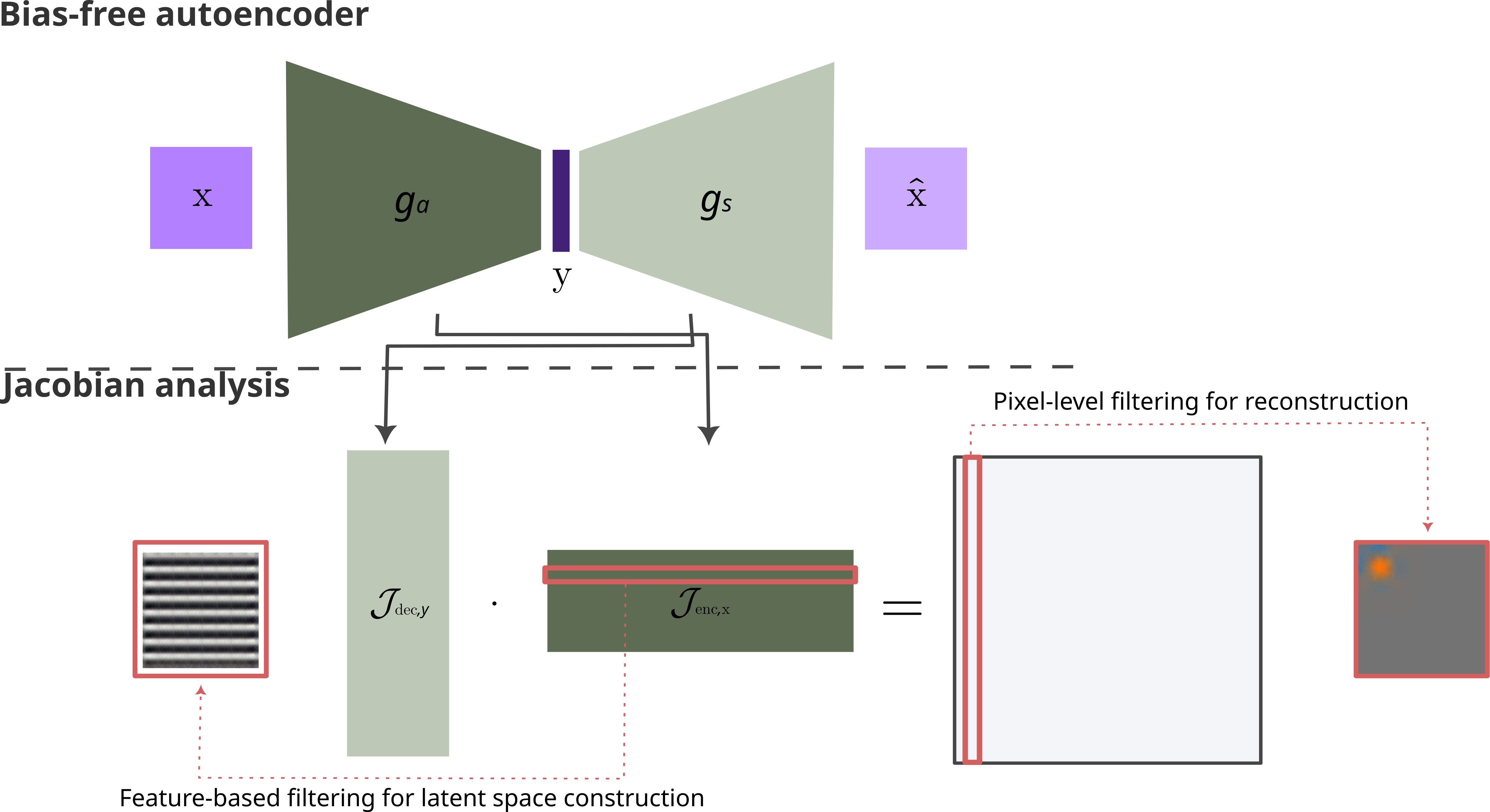}}
\caption{\small{\textbf{Bias-free autoencoders represented as two transformation matrices.} The matrices $\mathcal{J}_{\mathrm{enc},\vx}$ and $\mathcal{J}_{\mathrm{dec},\vy}$ are obtained from the Jacobians of the encoder $g_{\mathrm{enc}}(\cdot)$ and decoder $g_{\mathrm{dec}}(\cdot)$ evaluated at $\vx$ and $\vy$, respectively. They approximate the autoencoder's operations to obtain the reconstructed $\hat{\vx}$. The rows of $\mathcal{J}_{\mathrm{enc},\vx}$ are interpreted as learned encoder filters used to construct the latent space, while the columns of the resulting matrix product represent pixel-level filtering operations used for image reconstruction (examples in Figure~\ref{fig:individual_comb_filters})}.}
\label{fig:diagram}
\end{figure}

With these Jacobians, we can interpret the filtering process, expressing the knowledge from deep neural networks as two projection matrices $\mathcal{J}_{\mathrm{enc},\vx} $ and $\mathcal{J}_{\mathrm{dec},\vy}$ (Figure~\ref{fig:diagram}). This formulation enables pixel-wise analysis of the filtering process, as illustrated in Figure~\ref{fig:individual_comb_filters}. Since this filtering process is inherently adaptive, we investigate \textit{how similar these projections are across different images}, and whether there are \textit{shared components in the filtering behavior that are consistent across inputs} (Section~\ref{sec:generalization}). Finally, if these models show some generalization across inputs, \textit{we can investigate the possibility of deducing low-complexity filtering from complex models}~(Section~\ref{sec:comp_3ch}).

\vspace{-5mm}

\section{Generalization across images}
\label{sec:generalization}
We investigate whether the autoencoder filtering operations are correlated across images.
This helps determine whether the autoencoder behaves, at least partially, like a transform that decomposes images into shared components. We examine: (i) which latent features of $\vy$ have the greatest impact on performance across patches $\mathcal{P}$; (ii) how many features in $\vy$ are sufficient to reconstruct the original images with minimal loss; (iii) whether the encoder Jacobian row $\mathcal{J}_{\mathrm{enc},\vx,(i,.)}$ associated with a feature $i$ are correlated across $\mathcal{P}$; and, (iv) whether the most important features are also the most correlated across images. If a feature $i$ has Jacobian rows $\mathcal{J}_{\mathrm{enc},\vx,(i,.)}$ that are highly correlated across images, (v) we want to evaluate the impact of using a single Jacobian row (referred to as filter $i$) for all images to extract the corresponding feature. 
\vspace{-3mm}


\subsection{Finding important features}
\label{sec:important}
\begin{figure}[!ht]
    \centering
    \begin{subfigure}{0.45\linewidth}
        \centering
        \includegraphics[width=\linewidth]{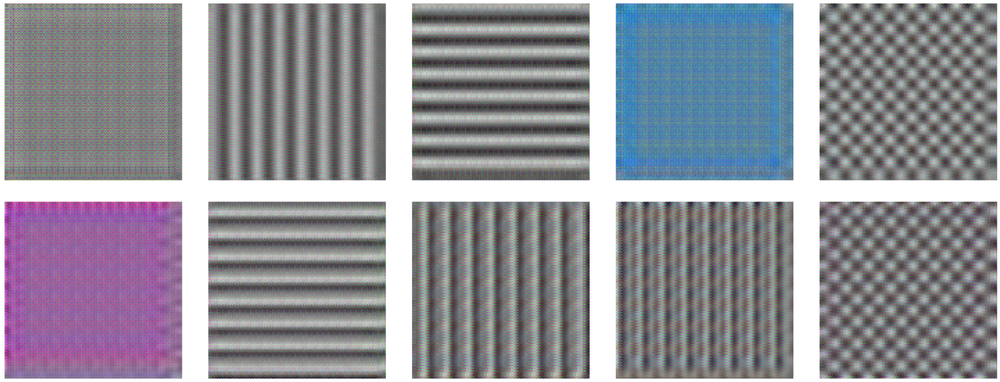}
        \caption{$\lambda=0.05$}
    \end{subfigure}
    ~
    \begin{subfigure}{0.45\linewidth}
        \centering
        \includegraphics[width=\linewidth]{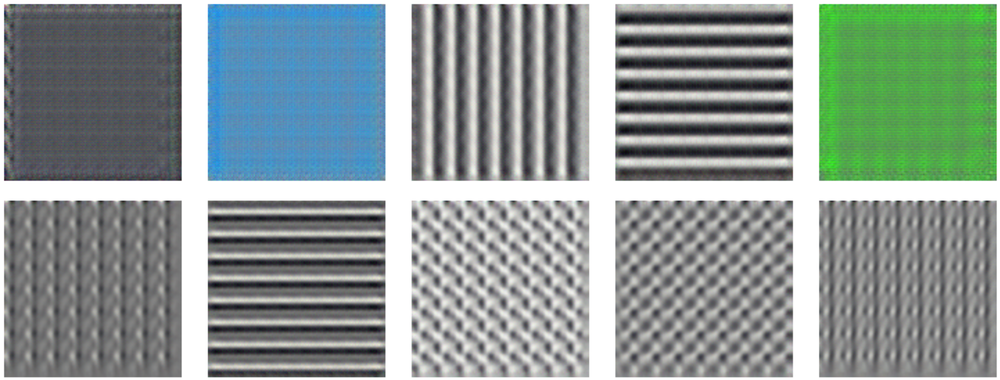}
        \caption{$\lambda=2.0$}
    \end{subfigure}
    \caption{\small{\textbf{Top-10 filters encoders.} We show the obtained encoder filters for two different compression models. In (a), high compression model with $\lambda=0.05$. In (b), low distortion model with $\lambda=2.0$, this model is used to generate Figure~\ref{fig:individual_comb_filters}, with filtering operations per pixel.}}
    \label{fig:filters}
\end{figure}

Different from transforms such as the $\operatorname{DCT}$, which order components by importance in signal reconstruction, the features from $\vy$ are not ranked by importance.  However, the magnitude of each row of the Jacobian $\mathcal{J}_{\mathrm{enc},\vx}$ from Equation~\eqref{eq:bf_encoder} (\textit{i.e.}, each filter) can be interpreted as a measure of the influence of the input features on a specific output feature $\vy_i$, as captured by the gradient magnitude. When combined with the magnitude of the corresponding $\vy_i$ feature value, this provides an indication of both the strength of the feature and the extent to which it is influenced by the input. This idea is inspired by post-hoc explainability methods~\cite{selvaraju:2017:gradcam, sundararajan:2017:gradAct}. Therefore, we define a score that combines the gradient magnitude with the feature value, to rank the filters by importance $s_{\mathbf{x},(i)} = \left\| a_{\mathbf{x},(i)} \right\| \cdot |\vy_i|$,
 where $a_{\vx,(i)} =\left[\mathcal{J}_{\mathrm{enc},\vx,(i,1)},\ldots,\mathcal{J}_{\mathrm{enc},\vx,(i,N)}\right]$ is the $i$-th corresponding row in $\mathcal{J}_{\mathrm{enc},\vx}$, which can be interpreted as an image-adaptive filter and visualized in a way similar to Figure~\ref{fig:filters}. The score $s_{\mathbf{x},(i)}$ is used to rank features according to their importance in reconstructing $\mathbf{x}$. We average this score over the patches $\mathcal{P}$, to obtain a global importance measure:
\begin{equation}
s_i = \frac{1}{|\mathcal{P}|} \sum_{\mathbf{x} \in \mathcal{P}} s_{\mathbf{x},(i)}.
\label{eq:filter_score_avg}
\end{equation}
\vspace{-3mm}
\begin{figure}[!ht]
\centering
\centerline{\includegraphics[width=0.8\linewidth]{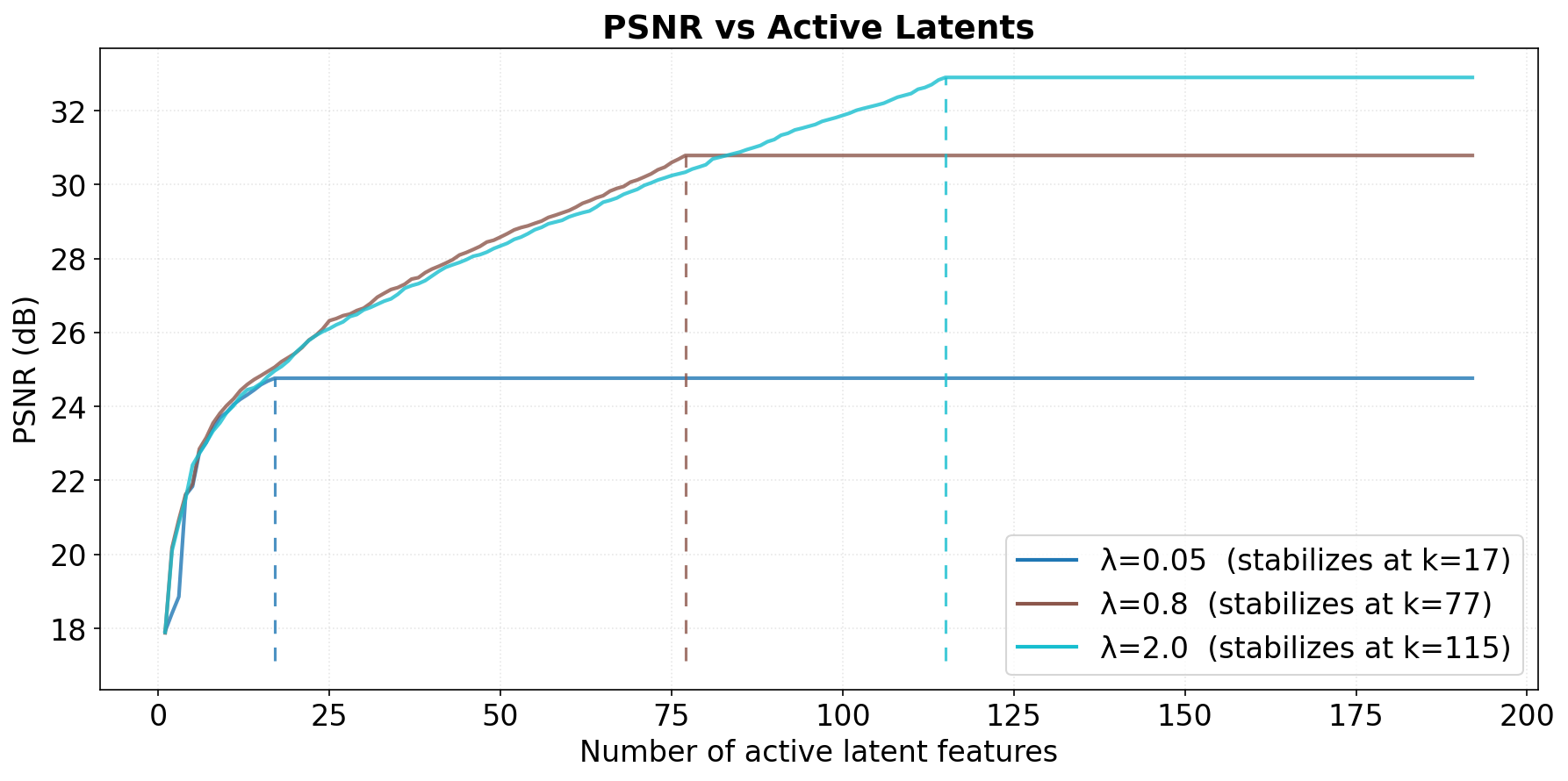}}
\caption{\small{\textbf{Reconstruction quality for features inclusion and different $\lambda$.} To reduce reconstruction distortion (i.e., by increasing $\lambda$), the models learn more informative features, as expected. However, out of the 192-dimensional feature space, none of the models uses more than two-thirds of the available dimensions, with the highest compression rate model using fewer than 25 effective features.}}
\label{fig:svd}
\end{figure}
\textbf{Sufficient features.} Using the proposed feature-importance score, we expect the reconstruction quality to increase smoothly as features are iteratively included from the most to the least important. Such behavior would indicate that features are being selected in correct order. To reduce unnecessary computation, we also analyze the point at which the reconstruction curve stabilizes in terms of PSNR, allowing us to retain only the features that are sufficient for reconstruction quality close to the maximum obtained by the model. Figure~\ref{fig:svd} presents the reconstruction curves for the three models under different values of $\lambda$, confirming the effectiveness of the proposed order. The results show an increasing concentration of reconstruction energy as a function of $\lambda$ (rate-distortion trade-off), suggesting that this parameter directly controls the amount of information transmitted, as expected in the compression pipeline. Based on the stabilization point (\textcolor{black}{Appendix~B}) of each curve, we retain only $k^{*}=17$, $77$, and $115$ features for $\lambda=0.05$, $0.8$, and $2.0$, respectively.

\vspace{-5mm}

\subsection{Features importance and cross-image correlation}
As previously discussed, we aim to verify whether the ordering of the most important filters, induced by the scores in Equation~\eqref{eq:filter_score_avg} corresponds to highly correlated filters across different images. In particular, we investigate whether the most important Jacobian rows (i.e., those capturing the largest amount of energy) exhibit consistent structure across images. To this end, we analyze $p \times p$ patches, with $p$ chosen to match the receptive field of the encoder in the adopted architecture. \textcolor{black}{This allows us to isolate local filtering behavior and reduce the effect of spatial translations}. We create a dataset of patches \(\mathcal{P}\) and perform a correlation analysis on the Jacobians.

For each feature $i$, let $\mathcal{J}_{\mathrm{enc},\vx,(i,.)}^{(n)} \in \mathbb{R}^{d}$ denote the $i$-th row of the Jacobian matrix computed for the $n$-th image patch $\mathbf{x}^{(n)}$. \textcolor{black}{Since our goal is to compare the frequency characteristics of the resulting Jacobian filters rather than their spatial alignment}, each Jacobian row is transformed into the frequency domain using the Fourier transform, obtaining $\mathbf{A}_i^{(n)}$. The correlation score for feature $c_i$ is then defined as the mean pairwise Pearson correlation between the flattened magnitude spectra across all $\binom{|\mathcal{P}|}{2} = \frac{|\mathcal{P}|(|\mathcal{P}|-1)}{2}$ patch pairs:
\begin{equation}
    c_i = \frac{2}{|\mathcal{P}|(|\mathcal{P}|-1)}
          \sum_{\substack{n_1,n_2 \,\in\, \mathcal{P} \\ n_1 < n_2 <|\mathcal{P}|}}
          \rho\!\left(\mathbf{A}_i^{(n_1)},\, \mathbf{A}_i^{(n_2)}\right)
    \label{eq:correlation_score}
\end{equation}
where $\rho(\mathbf{u}, \mathbf{v})$ denotes the Pearson correlation coefficient between two flattened vectors $\mathbf{u}, \mathbf{v} \in \mathbb{R}^{d}$. Given the importance score $s_i$ from Equation~\eqref{eq:filter_score_avg} and the correlation score $c_i$ from Equation~\eqref{eq:correlation_score}, we define two rankings as the $\operatorname{argsort}$ in descending order, $\mathcal{R}^{s} = \operatorname{argsort}_{i}(s_i)_{\downarrow}$ and  $\mathcal{R}^{c} = \operatorname{argsort}_{i}(c_i)_{\downarrow}$, 
such that $\mathcal{R}^{s}(r)$ and $\mathcal{R}^{c}(r)$ denote the feature assigned to rank $r \in \{1,\ldots,|\vy|\}$ under each criterion, where $|\vy|$ is the size of the latent space.
We further define the top-$k$ sets as $\mathcal{T}^{s}_{k} = \{\mathcal{R}^{s}(1),\ldots,\mathcal{R}^{s}(k)\}$ and $\mathcal{T}^{c}_{k} = \{\mathcal{R}^{c}(1),\ldots,\mathcal{R}^{c}(k)\}$. To assess whether the most energetic features are also the most correlated across images, we evaluate the agreement between $\mathcal{R}^{s}$ and $\mathcal{R}^{c}$ using two complementary metrics: Normalized Discounted Cumulative Gain (NDCG@$k$)~\cite{jarvelin:2002:ndcg} and Kendall's tau ranking correlation~\cite{kendall:1948:rank}. NDCG measures the correspondence between two rankings by assigning greater importance to agreement at higher-ranked positions, whereas Kendall's tau evaluates whether feature pairs preserve the same relative order within the top $k$. All two metrics are evaluated as a function of the cutoff $k$, assessing whether the agreement between the two rankings is at the top features or distributed across the full set.
We test $434$ images randomly sampled ($0.1\%$ of the training dataset) and compute the Jacobians on their central crops. We use $16 \times 16$ patches. For a network with $192$ latents (\textcolor{black}{details in Appendix A}), we compute the scores from Equation~\eqref{eq:filter_score_avg} for $i \in [1,192]$ , using a dataset of size $|\mathcal{P}| = 434$.
\vspace{-3mm}
\begin{figure}[!ht]
\centerline{\includegraphics[width=\linewidth]{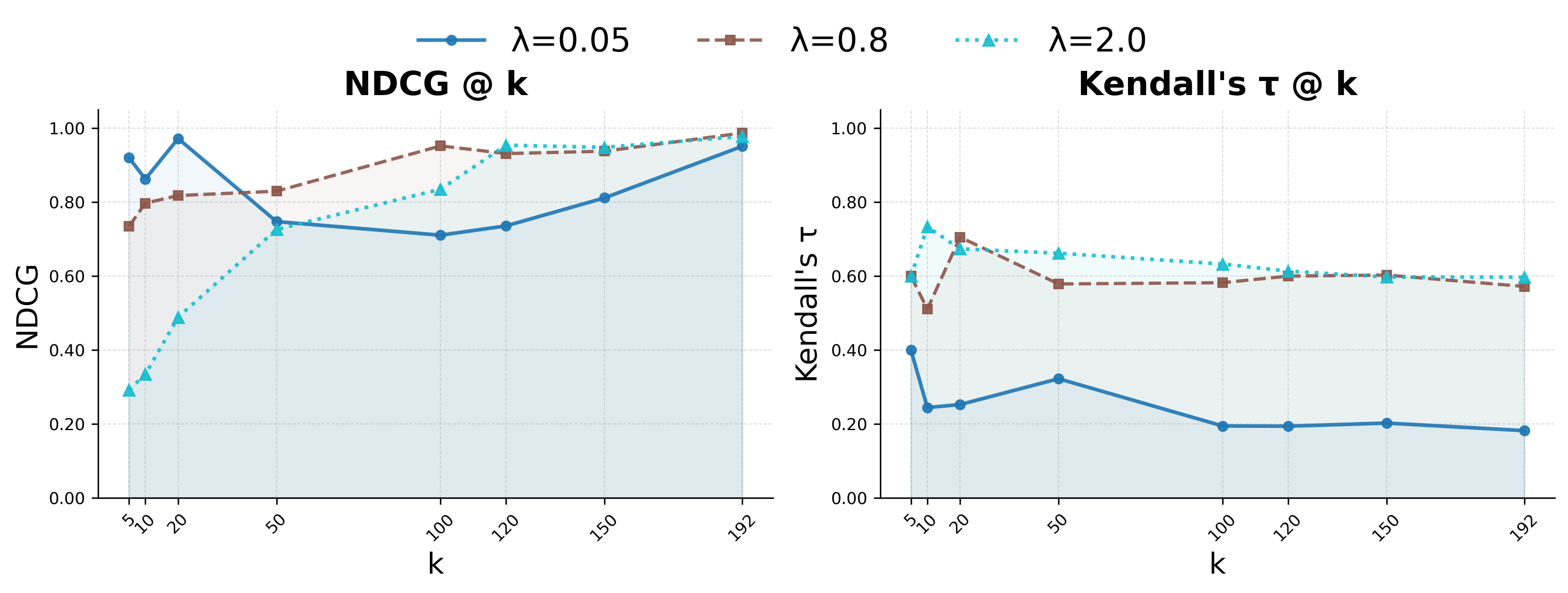}}
\caption{\small{\textbf{Ranking agreement metrics for $\lambda$.} Increasing NDCG values confirm agreement between the rankings $\mathcal{R}^{s}$ (feature importance) and $\mathcal{R}^{c}$ (correlation across images). This agreement stabilizes near the number of important features $k^*$. Kendall's $\tau$ shows higher agreement among top features and gradually decreases, indicating that the most important features are also the most correlated across images.}}
\label{fig:energy_correlation}
\end{figure}

\textbf{Importance vs. correlation results. }The metrics presented in Figure~\ref{fig:energy_correlation} NDCG) confirm that $\mathcal{R}^{s}$ and $\mathcal{R}^{c}$ are correlated in how they order the important filters: features with high cross-image correlation in the frequency domain tend to also present the highest Jacobian energy, supporting the hypothesis that energetic and correlated filters are mostly the same. This holds approximately up to the threshold $k^{*}= 17,77, 115$ identified in Section~\ref{sec:important}, beyond which correlation degrades, confirming $k^{*}$ as a threshold between discriminative and redundant features. \textcolor{black}{Beyond this threshold, features may behave more randomly as they no longer provide meaningful information for reconstruction (reflecting on the decreasing curves). Our main interest is in the initial slope of the curves: a steeper increase indicates that agreement between rankings grows faster among the top-ranked features.} Moreover, Kendall's $\tau$ (which measures pairwise concordance between rankings) confirms that the strongest correspondence between $\mathcal{R}^{s}$ and $\mathcal{R}^{c}$ is concentrated at the highest-ranked features, \textit{i.e.} the most energetic and correlated filters coincide, with concordance progressively weakening as $k$ increases towards $k^{*}$.
This behavior is consistent with the intuition from handcrafted transform coding: classical transforms such as the $\operatorname{DCT}$ concentrates signal energy into a few low-frequency coefficients, which are simultaneously the most energetic, and the most consistent across natural images. The learned encoder $g_{\mathrm{enc}}(\cdot)$ appears to replicate this, the cross-image most energetic filters are highly correlated across images, which indicates we can combine the top filters to use them statically.

\vspace{-5mm}

\subsection{Combining multiple Jacobians}
\label{sec:combination}
If the filters $i$ ($\mathcal{J}_{\mathrm{enc},\vx,(i,.)}^{(m)}$ for an image $m \in \mathcal{D}$) ranked in the top-$k$ of $\mathcal{R}^{s}$ are highly correlated, then they may be transferable across images. In particular, it should be possible to use the most correlated filters from one image to process the other with minimal loss. To evaluate this hypothesis, we compute an averaged filter representation over a subset of image Jacobians, obtaining an average Jacobian matrix $ J^*_{enc} = \frac{1}{|\mathcal{D}|} \sum_{m \in \mathcal{D}} \mathcal{J}_{\mathrm{enc}}^{(m)}$.  We can repeat the process with $J_{dec}$ for obtaining $J^*_{dec} = \frac{1}{|\mathcal{D}|} \sum_{m \in \mathcal{D}} \mathcal{J}_{\mathrm{dec}}^{(m)}$ (average Jacobian decoder), with $\mathcal{D}$ the subset of images with a fixed dimension.

\textbf{Filters visualization.} We present the average top-10 filters obtained from $J^*_{enc}$, visualized as the input image dimensions $(128,128,3)$ in the  Figure~\ref{fig:filters}. 
The encoder filters show the nature of the filtered signal, and even across models with different $\lambda$ values, they show strong similarities. In particular, the most important filters tend to emphasize vertical and horizontal structures, as well as color-related characteristics. 

\vspace{-2mm}

\section{Rate-distortion trade-off}
\label{sec:comp_3ch}
Using $J^*_{enc}$ and $J^*_{dec}$, we obtain two linear projection matrices from a nonlinear model, which are used for image compression and reconstruction. We compare it with the original network, as well as with the KLT and DCT methods (details in Appendix~C). 
We evaluate models in Kodak dataset~\cite{kodak:1999:Online} (\textcolor{black}{not seen in training}), trained with $\lambda = 2.0$ (details in Appendix~A). \textcolor{black}{For a fair comparison, all methods are evaluated using the same, simpler probability-mass entropy model (Appendix~D), avoiding any advantage for the neural network and Jacobian-based methods over DCT and KLT.} We show results from the model with $\lambda=2.0$ (better reconstructions).  We show extra comparison in a more realistic scenario in Appendix~K.
\vspace{-3mm}
\begin{figure}[!ht]
\centerline{\includegraphics[width=0.8\linewidth]{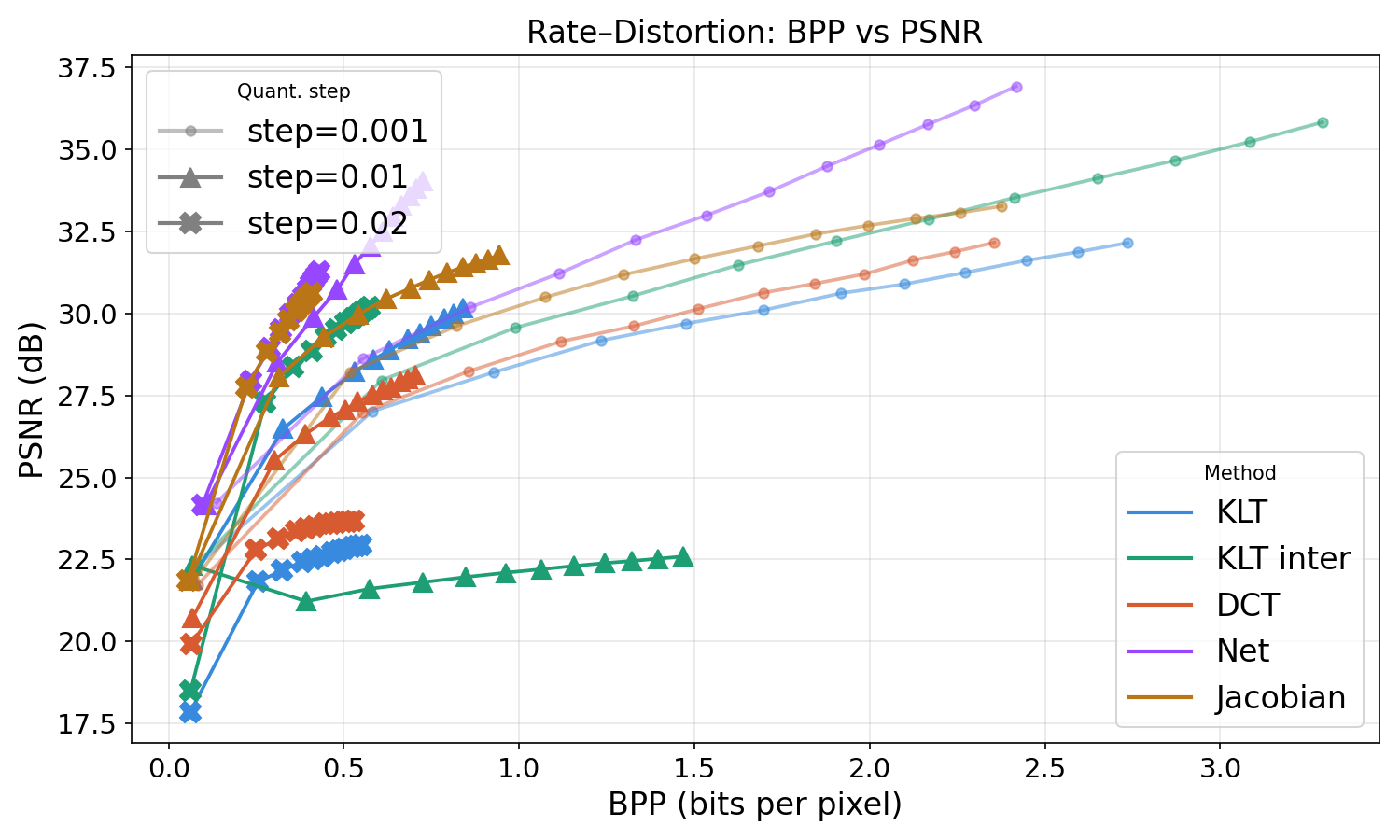}}
\caption{\small{\textbf{Comparison of quantization steps and increasing features.} We progressively increase the number of features according to their importance. We evaluate the original model, the Jacobian-based approach, and the transform-based methods DCT and KLT, considering separate and joint channel analysis (inter-channel) for KLT. Our Jacobian filtering method outperforms the transform-based and achieves PSNR values comparable to the original model, particularly at low BPP rates. \textcolor{black}{Similar trends for CLIC dataset (Appendix~I).}}}
\label{fig:quantization_features}
\end{figure}

\textbf{Comparing compression methods.} 
In Figure~\ref{fig:quantization_features}, we vary $q$ and the number of features (ordered from most to least energetic) used in the reconstruction across all evaluated models. We limit the analysis to 115 sufficient features (refer to Figure~\ref{fig:svd}). We select the model with $\lambda=2.0$, as it yields the best reconstruction performance. Figure~\ref{fig:quantization_features} presents results for $q \in \{0.001, 0.1, 0.2\}$ while also varying the number of features included in the reconstruction. Each curve starts with a single feature (leftmost point) and progressively increases up to 115 features, which account for most of the explained variance in this model (see Figure~\ref{fig:svd}). 
Each feature corresponds to a block of $8\times8$ coefficients, \textcolor{black}{since the encoder downsamples $128\times128$ images by 16  (for the Net and Jacobian).
For the other transforms, one feature corresponds to one coefficient from each of the 64 blocks of $16\times16$ resulting in a total of 64 coefficients. Therefore, all methods transmit the same number of coefficients for any $k$.}


We observe that the Jacobian behaves similarly to the network, particularly at low BPP rates. Across all quantization steps, the Jacobian achieves a good PSNR–BPP trade-off. Second, both Jacobian and  Net, in contrast to conventional transforms such as the DCT and KLT, maintain a stronger PSNR–BPP trade-off at higher quantization steps. In other words, they can further reduce BPP compared to the other methods while preserving relatively high PSNR, even with the simpler entropy model used. This may help explain the improved performance of learning-based models in compression tasks, as they seem to learn decompositions that better account for rate–distortion trade-offs, unlike classical transforms that focus on energy compaction and decorrelation. \textcolor{black}{Table~\ref{tab:complexity_encdec} shows that, although less efficient than DCT and KLT, the Jacobian achieves up to $91\%$ fewer MACs/pixel than the original model.}
\vspace{-3mm}
\begin{table}[!ht]
\centering
\caption{\textcolor{black}{\small{Encoder + decoder MACs/pixel and reduction relative to the bias-free factorized-prior network. $k$: retained latent features.}}}
\label{tab:complexity_encdec}
\footnotesize
\setlength{\tabcolsep}{3pt}
\begin{tabular}{@{}lrrrrrr@{}}
\toprule
 & \multicolumn{2}{c}{$k=115$} & \multicolumn{2}{c}{$k=77$} & \multicolumn{2}{c}{$k=17$} \\
\cmidrule(lr){2-3}\cmidrule(lr){4-5}\cmidrule(l){6-7}
Method & MACs & Red. ($\%$) & MACs & Red. ($\%$) & MACs & Red. ($\%$) \\
\midrule
Network         & 73{,}600 & --    & 73{,}600 & --    & 73{,}600 & --    \\
Jacobian        & 44{,}160 & 40.00 & 29{,}568 & 59.83 &  6{,}528 & 91.13 \\
KLT             &      248 & 99.66 &      172 & 99.77 &       52 & 99.93 \\
DCT$^{\dagger}$ &      210 & 99.71 &      210 & 99.71 &      210 & 99.71 \\
\bottomrule
\end{tabular}

\vspace{1pt}
{\scriptsize $^{\dagger}$Full separable $16\times16$ DCT/IDCT, independent of $k$.}
\end{table}

\vspace{-5mm}

\section{Conclusion}
Deep learning-based image compression models face practical limitations due to their size and computational cost. In this work, we study bias-free autoencoders for the image compression task, to investigate whether universal structures exist that could help simplify these data-adaptive models. We observe that a subset of encoder operations is shared across images (highly correlated) and can be interpreted as image filters. Moreover, the filters that are most consistent across images are also among those that contribute most to reconstruction energy. Using the average filters across images for both encoding and decoding presents results close to the original model, particularly at low bit rates when only the most energetic components are retained. A comparison with classical transform methods such as the DCT and KLT shows that these learned filters benefit from data adaptation inherited from the original models, achieving a better rate–distortion trade-off. \textcolor{black}{However, our analysis is limited to bias-free ReLU-based architectures and depends on the strategy used to combine the Jacobians. We plan to extend to more general models and to other combination strategies.}



\balance
\bibliographystyle{IEEEtran}
\bibliography{main}

\begin{thebibliography}{10}
\providecommand{\url}[1]{#1}
\csname url@samestyle\endcsname
\providecommand{\newblock}{\relax}
\providecommand{\bibinfo}[2]{#2}
\providecommand{\BIBentrySTDinterwordspacing}{\spaceskip=0pt\relax}
\providecommand{\BIBentryALTinterwordstretchfactor}{4}
\providecommand{\BIBentryALTinterwordspacing}{\spaceskip=\fontdimen2\font plus
\BIBentryALTinterwordstretchfactor\fontdimen3\font minus
  \fontdimen4\font\relax}
\providecommand{\BIBforeignlanguage}[2]{{%
\expandafter\ifx\csname l@#1\endcsname\relax
\typeout{** WARNING: IEEEtran.bst: No hyphenation pattern has been}%
\typeout{** loaded for the language `#1'. Using the pattern for}%
\typeout{** the default language instead.}%
\else
\language=\csname l@#1\endcsname
\fi
#2}}
\providecommand{\BIBdecl}{\relax}
\BIBdecl

\bibitem{balle:iclr:2017:factorized}
J.~Ballé, V.~Laparra, and E.~P. Simoncelli, ``End-to-end optimized image
  compression,'' in \emph{5th International Conference on Learning
  Representations (ICLR)}, 2017, pp. 1--27.

\bibitem{balle:iclr:2018:hyper}
J.~Ballé, D.~Minnen, S.~Singh, S.~J. Hwang, and N.~Johnston, ``Variational
  image compression with a scale hyperprior,'' in \emph{6th International
  Conference on Learning Representations (ICLR)}, 2018, pp. 1--23.

\bibitem{yang:2023:trends:intro1}
Y.~Yang, S.~Mandt, and L.~Theis, ``An introduction to neural data
  compression,'' \emph{Foundations and Trends in Computer Graphics and Vision},
  vol.~5, pp. 113--200, 2023.

\bibitem{Gilbert:2024:ahn}
M.~Gilbert, M.~Campos, and M.~Campista, ``Asymmetric autoencoders: An nn
  alternative for resource-constrained devices in iot networks,'' \emph{Ad Hoc
  Networks}, vol. 156, 02 2024.

\bibitem{yang:iccv:2023:shallow:decoder}
Y.~Yang and S.~Mandt, ``Computationally-efficient neural image compression with
  shallow decoders,'' in \emph{19th International Conference on Computer Vision
  (ICCV)}, 2023, pp. 1--23.

\bibitem{liang:2020:itj}
F.~Liang, W.~Yu, X.~Liu, D.~Griffith, and N.~Golmie, ``Toward edge-based deep
  learning in industrial internet of things,'' \emph{IEEE Internet of Things
  Journal}, vol.~7, no.~5, pp. 4329--4341, 2020.

\bibitem{hassija:2023:interpreting}
V.~Hassija, V.~Chamola, A.~Mahapatra, A.~Singal, D.~Goel, K.~Huang,
  S.~Scardapane, I.~Spinelli, M.~Mahmud, and A.~Hussain, ``Interpreting
  black-box models: A review on explainable artificial intelligence,''
  \emph{Cognitive Computation}, vol.~16, pp. 45--74, 2024.

\bibitem{mohan:2020:iclr}
S.~Mohan, Z.~Kadkhodaie, E.~P. Simoncelli, and C.~Fernandez-Granda, ``Robust
  and interpretable blind image denoising via bias-free convolutional neural
  networks,'' in \emph{8th International Conference on Learning Representations
  (ICLR)}, 2020, pp. 1--22.

\bibitem{kadkhodaie:2024:iclr}
Z.~Kadkhodaie, F.~Guth, E.~P. Simoncelli, and S.~Mallat, ``Generalization in
  diffusion models arises from geometry-adaptive harmonic representations,'' in
  \emph{International Conference on Learning Representations (ICLR)}, 2024.

\bibitem{lundberg:2017:shap}
S.~M. Lundberg and S.-I. Lee, ``A unified approach to interpreting model
  predictions,'' in \emph{Advances in Neural Information Processing Systems
  (NeurIPS)}, 2017, pp. 4765--4774.

\bibitem{ribeiro:2016:lime}
M.~T. Ribeiro, S.~Singh, and C.~Guestrin, ``“why should i trust you?”:
  Explaining the predictions of any classifier,'' in \emph{22nd ACM SIGKDD
  International Conference on Knowledge Discovery and Data Mining}.\hskip 1em
  plus 0.5em minus 0.4em\relax ACM, 2016, pp. 1135--1144.

\bibitem{RODRIGUES:2024:prl}
\BIBentryALTinterwordspacing
C.~M. Rodrigues, N.~Boutry, and L.~Najman, ``Transforming gradient-based
  techniques into interpretable methods,'' \emph{Pattern Recognition Letters},
  vol. 184, pp. 66--73, 2024. [Online]. Available:
  \url{https://www.sciencedirect.com/science/article/pii/S0167865524001764}
\BIBentrySTDinterwordspacing

\bibitem{ghorbani2019ace}
A.~Ghorbani, J.~Wexler, J.~Zou, and B.~Kim, ``Towards automatic concept-based
  explanations,'' in \emph{Advances in Neural Information Processing Systems
  (NeurIPS)}, 2019, pp. 9273--9282.

\bibitem{RODRIGUES:2024:infoSciences}
\BIBentryALTinterwordspacing
C.~{Mazini Rodrigues}, N.~Boutry, and L.~Najman, ``Unsupervised discovery of
  interpretable visual concepts,'' \emph{Information Sciences}, vol. 661, p.
  120159, 2024. [Online]. Available:
  \url{https://www.sciencedirect.com/science/article/pii/S0020025524000720}
\BIBentrySTDinterwordspacing

\bibitem{rodrigues2026xaitrees}
\BIBentryALTinterwordspacing
C.~M. Rodrigues, N.~Boutry, and L.~Najman, ``Explaining with trees:
  Interpreting cnns using hierarchies,'' \emph{Transactions on Machine Learning
  Research}, 2026. [Online]. Available:
  \url{https://openreview.net/forum?id=zjyWZh5IiI}
\BIBentrySTDinterwordspacing

\bibitem{Jamil:2023:EAAI}
S.~Jamil, M.~J. Piran, M.~Rahman, and O.-J. Kwo, ``Learning-driven lossy image
  compression: A comprehensive survey,'' \emph{Engineering Applications of
  Artificial Intelligence}, vol. 123, pp. 1--17, 2023.

\bibitem{selvaraju:2017:gradcam}
R.~R. Selvaraju, M.~Cogswell, A.~Das, R.~Vedantam, D.~Parikh, and D.~Batra,
  ``Grad-cam: Visual explanations from deep networks via gradient-based
  localization,'' in \emph{IEEE International Conference on Computer Vision
  (ICCV)}, 2017, pp. 618--626.

\bibitem{sundararajan:2017:gradAct}
M.~Sundararajan, A.~Taly, and Q.~Yan, ``Axiomatic attribution for deep
  networks,'' in \emph{34th International Conference on Machine Learning
  (ICML)}, 2017, pp. 1--10.

\bibitem{jarvelin:2002:ndcg}
K.~J{\"a}rvelin and J.~Kek{\"a}l{\"a}inen, ``Cumulated gain-based evaluation of
  ir techniques,'' \emph{ACM Transactions on Information Systems}, vol.~20,
  no.~4, pp. 422--446, 2002.

\bibitem{kendall:1948:rank}
M.~G. Kendall, \emph{Rank Correlation Methods}, 1st~ed.\hskip 1em plus 0.5em
  minus 0.4em\relax Griffin, 1948.

\bibitem{kodak:1999:Online}
R.~Franzen, ``Kodak lossless true color image suite,''
  \url{https://r0k.us/graphics/kodak/}, 1999 (accessed June 26, 2026).

\bibitem{karhunen:1947:linear}
K.~Karhunen, ``{\"U}ber lineare methoden in der wahrscheinlichkeitsrechnung,''
  \emph{Annales Academiae Scientiarum Fennicae}, vol.~37, pp. 1--79, 1947.

\bibitem{loeve:1948:fonctions}
M.~Lo{\`e}ve, ``Fonctions al{\'e}atoires de second ordre,'' in \emph{Processus
  Stochastiques et Mouvement Brownien}.\hskip 1em plus 0.5em minus 0.4em\relax
  Hermann, 1948.

\bibitem{rao:1990:dct}
K.~R. Rao and P.~Yip, \emph{Discrete Cosine Transform: Algorithms, Advantages,
  Applications}.\hskip 1em plus 0.5em minus 0.4em\relax Academic Press, 1990.

\bibitem{xue:ijvc:2019:vimeo}
T.~Xue, B.~Chen, J.~Wu, D.~Wei, and W.~T. Freeman, ``Video enhancement with
  task-oriented flow,'' \emph{International Journal of Computer Vision (IJCV)},
  vol. 127, no.~8, pp. 1106--1125, 2019.

\end{thebibliography}
\nocite{karhunen:1947:linear}
\nocite{loeve:1948:fonctions}
\nocite{rao:1990:dct}
\nocite{xue:ijvc:2019:vimeo}

\appendix
\section{Supplementary Material}

\textbf{A. Original models:} \textcolor{black}{For the experiments, we adapt the Factorized prior architecture~\cite{balle:iclr:2017:factorized} (four encoder and four decoder layers, with 128 filters per layer except for the 192-channel latent layer). trained on the Vimeo-90k dataset~\cite{xue:ijvc:2019:vimeo} (89,800 video clips). Bias terms are removed, and the GDN layers are replaced with ReLU. We trained several models with different values of \(\lambda\), and report results for three of them: \(\lambda \in \{0.05, 0.8, 2.0\}\). We followed the training protocol and data splits provided by CompressAI. For the feature importance and correlation analyses, we used 0.1\% of the training data.}

\textbf{B. Selection of $k^\star$}: \textcolor{black}{Let $P(k)$ denote the mean PSNR when reconstructing images using the $k$ most important latent features. We define the stabilization point $k^\star$ as the smallest number of features for which the PSNR range over a sliding window of $W=10$ consecutive values is below $\Delta=0.05$. If no such point exists, we set $k^\star=K$.}

\textbf{C. Rate-distortion trade-off (baseline details):} The Karhunen-Lo\`eve transform (\textbf{KLT})~\cite{karhunen:1947:linear, loeve:1948:fonctions} is trained on the same dataset $\mathcal{D}$ used to compute the averaged Jacobians. We evaluate two variants: a per-channel KLT and an inter-channel KLT. In both cases, the transform is applied to image blocks of size $16 \times 16$ pixels. We apply the Discrete Cosine Transform (\textbf{DCT})~\cite{rao:1990:dct} on a per-channel basis using blocks of $16 \times 16$ pixels. The transmitted coefficients are selected in a zig-zag order, following the JPEG standard, within each block.


\textbf{D. Rate-distortion trade-off (Entropy model):} We test different quantization steps $q \in \{0.001, 0.1, 0.2\}$ and, we use the training set $\mathcal{D}$ to obtain a discrete probability mass function for the entropy evaluation. Let $z$ be the transform and normalized coefficients obtained from a given method applied to samples in $\mathcal{D}$. For a quantization step $q$, we define the quantized variable i as $\hat{z} = Q_q(z) = q \cdot \left\lfloor \frac{z}{q} \right\rceil$. From the quantized values $\hat{z}$ from samples in $\mathcal{D}$, we estimate a discrete probability mass function as:
\begin{equation}
    p(\hat{z} = kq) = \frac{1}{N} \sum_{i=1}^{N} \mathbf{1}\{\hat{z}_i = kq\},
    \label{eq:pmf}
\end{equation}
where $N$ is the total number of coefficients (or latent features) in the dataset $\mathcal{D}$, $k$ is the bin index, and $\mathbf{1}\{\cdot\}$ is the indicator function. The entropy is then computed by: 
\begin{equation}
   H(\hat{z}) = - \sum_{k} p(\hat{z} = kq)\,\log_2 p(\hat{z} = kq),
    \label{eq:entropy}
\end{equation}
and used to calculate bits per pixel (BPP) by normalizing with respect to the number of image pixels. We vary $q$ and the number of features (from the most to the least energetic) used in the reconstruction for all tested models. 

\begin{figure}[!hb]
\centering
\includegraphics[width=0.9\linewidth]{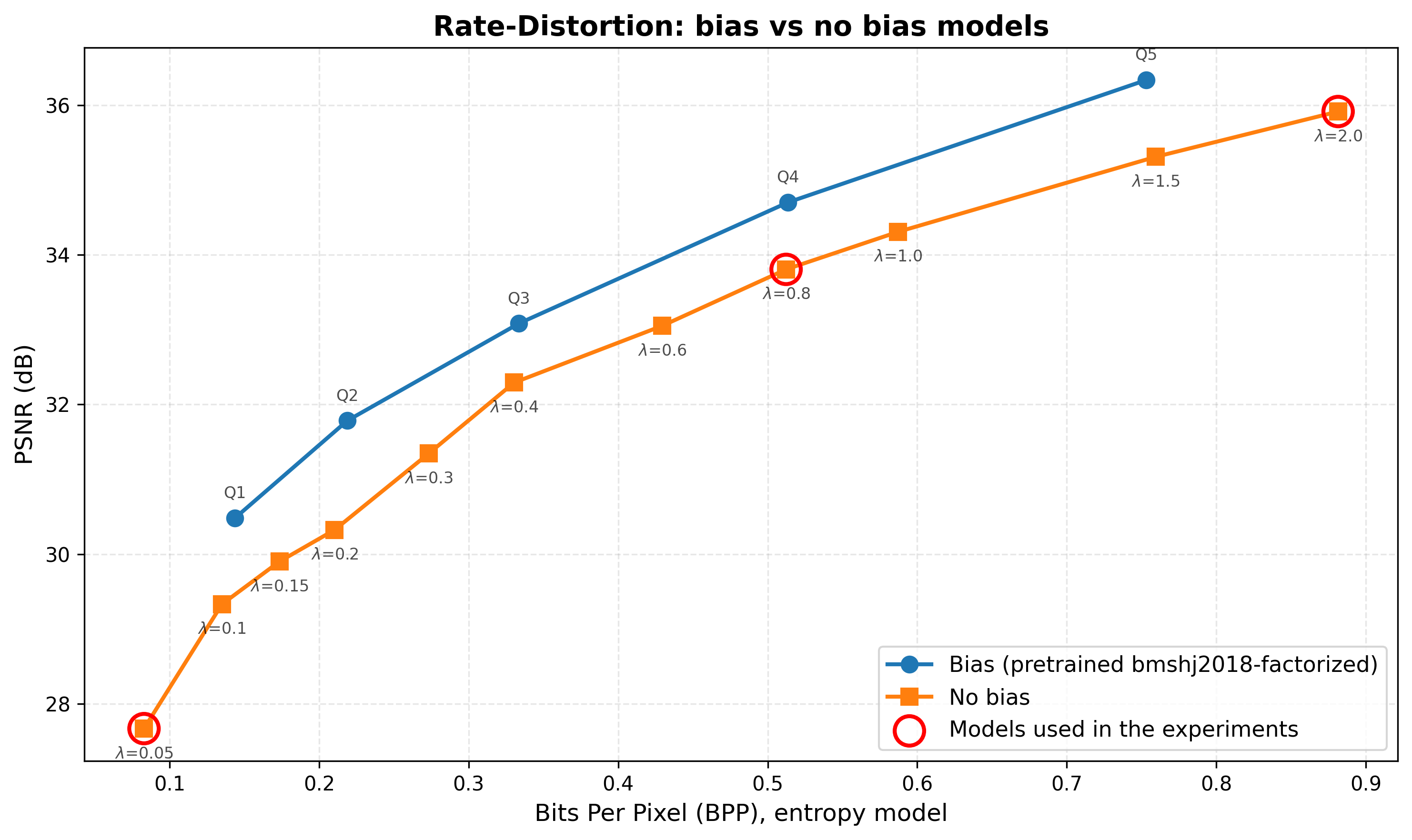}
\caption{\small{\textbf{Comparison original vs. bias-free.} Original architecture compared to the modified bias-free architecture using ReLU activations, across multiple models trained with different rate–distortion trade-offs.}}
\label{fig:bias_nobias}
\end{figure}

\textbf{E. Comparison original and bias-free models:} Fig.~\ref{fig:bias_nobias} compares the original architecture with the modified bias-free architecture using ReLU activations, across multiple models trained with different rate–distortion trade-offs. All models were trained on the same dataset to ensure a fair comparison. The results show that removing biases leads to a slight performance degradation, while preserving similar overall behavior across the different rate–distortion trade-offs. This performance loss is precisely one of the reasons we explicitly acknowledge the use of bias-free layers as a limitation of our approach.

\begin{figure*}[!ht]
    \centering
    \begin{subfigure}{0.8\linewidth}
        \centering
        \includegraphics[width=\linewidth]{3nets_psnr_only_my_method.png}
        \caption{PSNR vs.\ number of features.}
        \label{fig:psnr_main}
    \end{subfigure}

    \vspace{0.5em}

    \begin{subfigure}{0.48\linewidth}
        \centering
        \includegraphics[width=\linewidth]{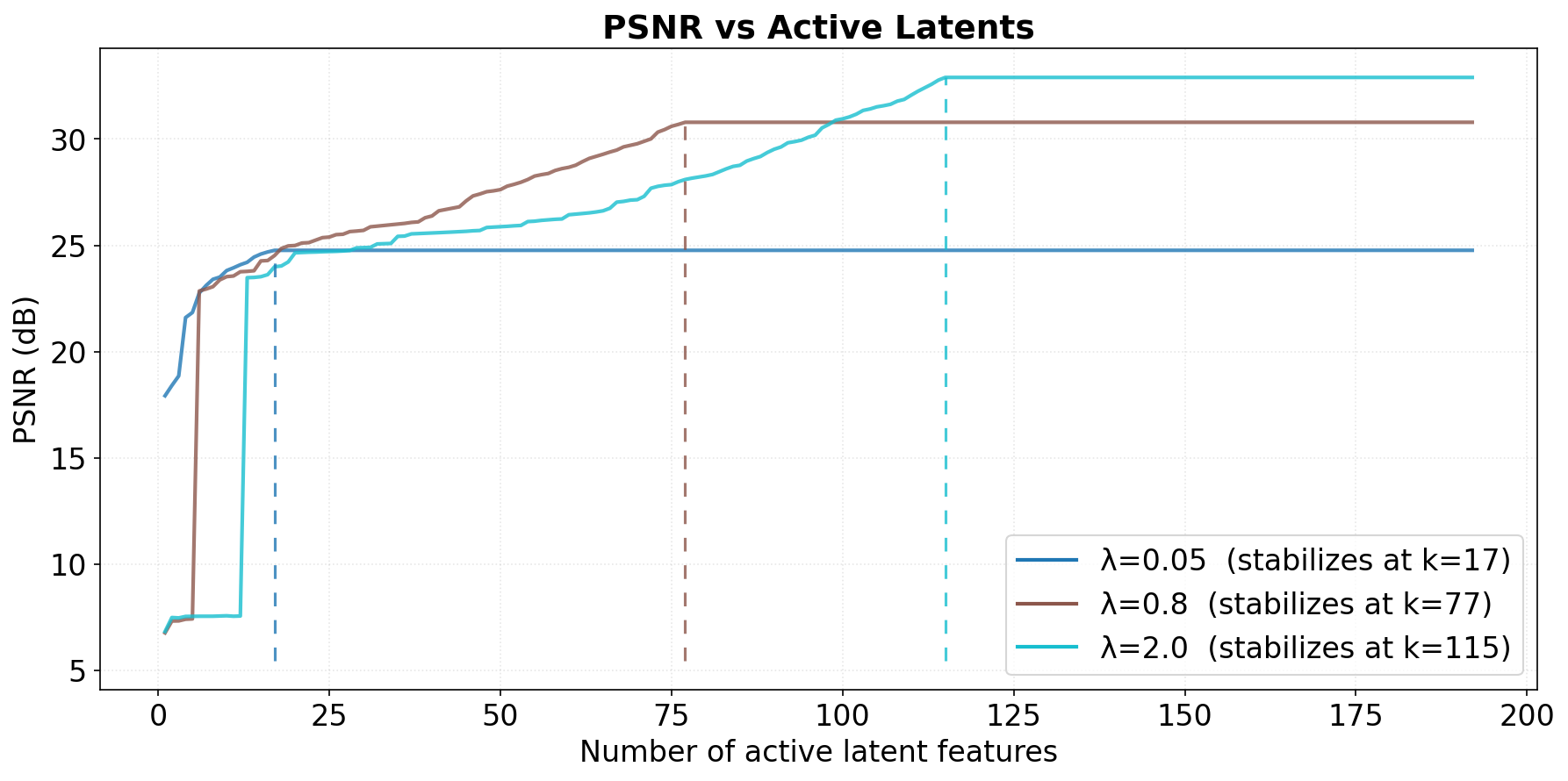}
        \caption{Energy-based ordering.}
        \label{fig:psnr_energy}
    \end{subfigure}
    \hfill
    \begin{subfigure}{0.48\linewidth}
        \centering
        \includegraphics[width=\linewidth]{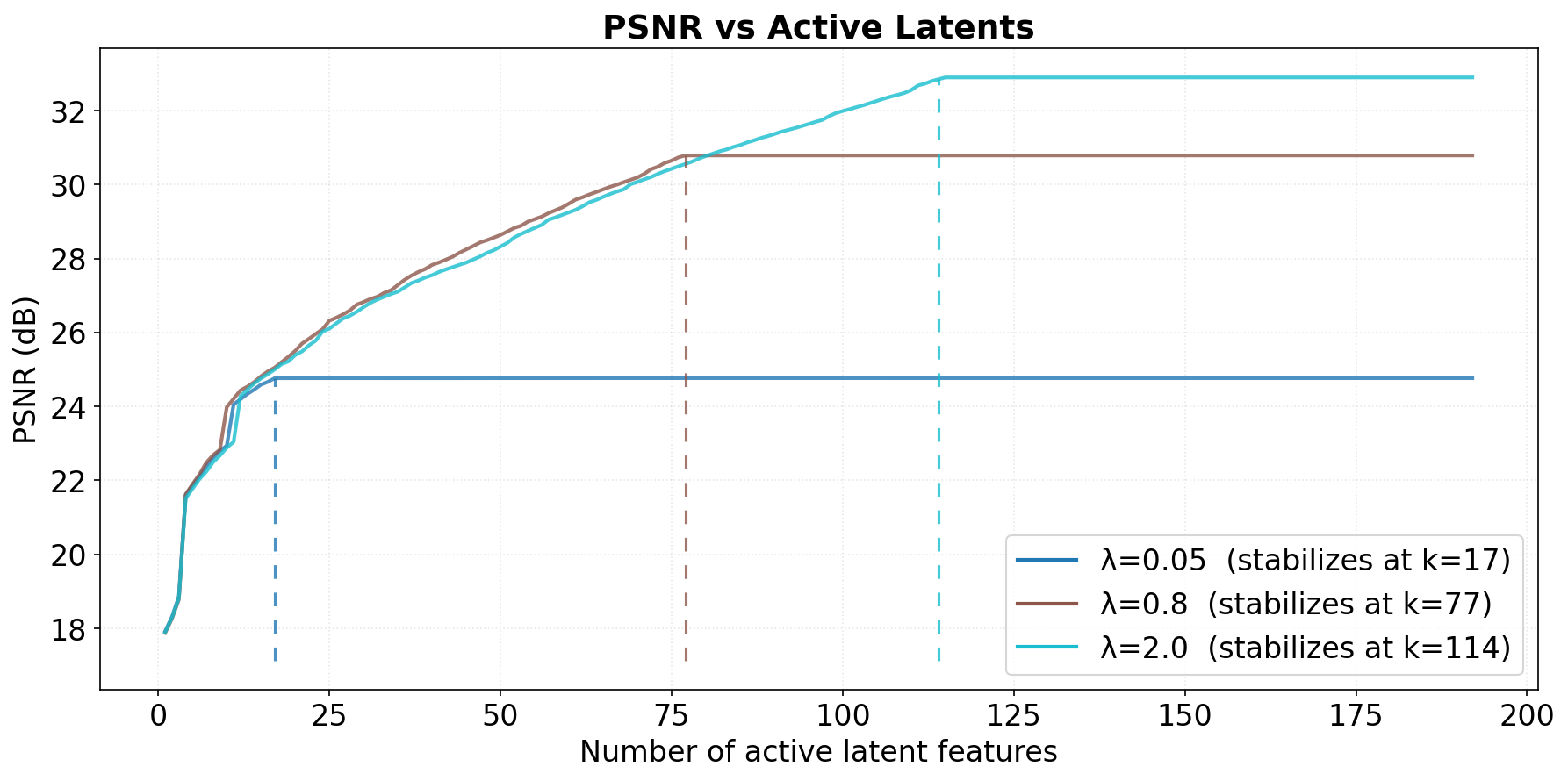}
        \caption{Activation-based ordering.}
        \label{fig:psnr_activation}
    \end{subfigure}

    \caption{\small{\textbf{Comparison of different ordering methods.} PSNR results of the proposed latent ordering methods for the three networks.}}
    \label{fig:psnr_3nets_ordering}
\end{figure*}

\textbf{F. Ablation on latents ordering:} We include an ablation study using two additional feature orderings: activation magnitude and gradient magnitude, considered separately. Fig.~\ref{fig:psnr_3nets_ordering} shows that combining the two metrics, as proposed in the paper, results in a smoother increase in reconstruction quality. This is expected, as a better ordering places the most informative features first, leading to higher reconstruction quality when only a few features are used and a faster, smoother increase as more are added.

\begin{figure}[!hb]
\centering
\includegraphics[width=0.9\linewidth]{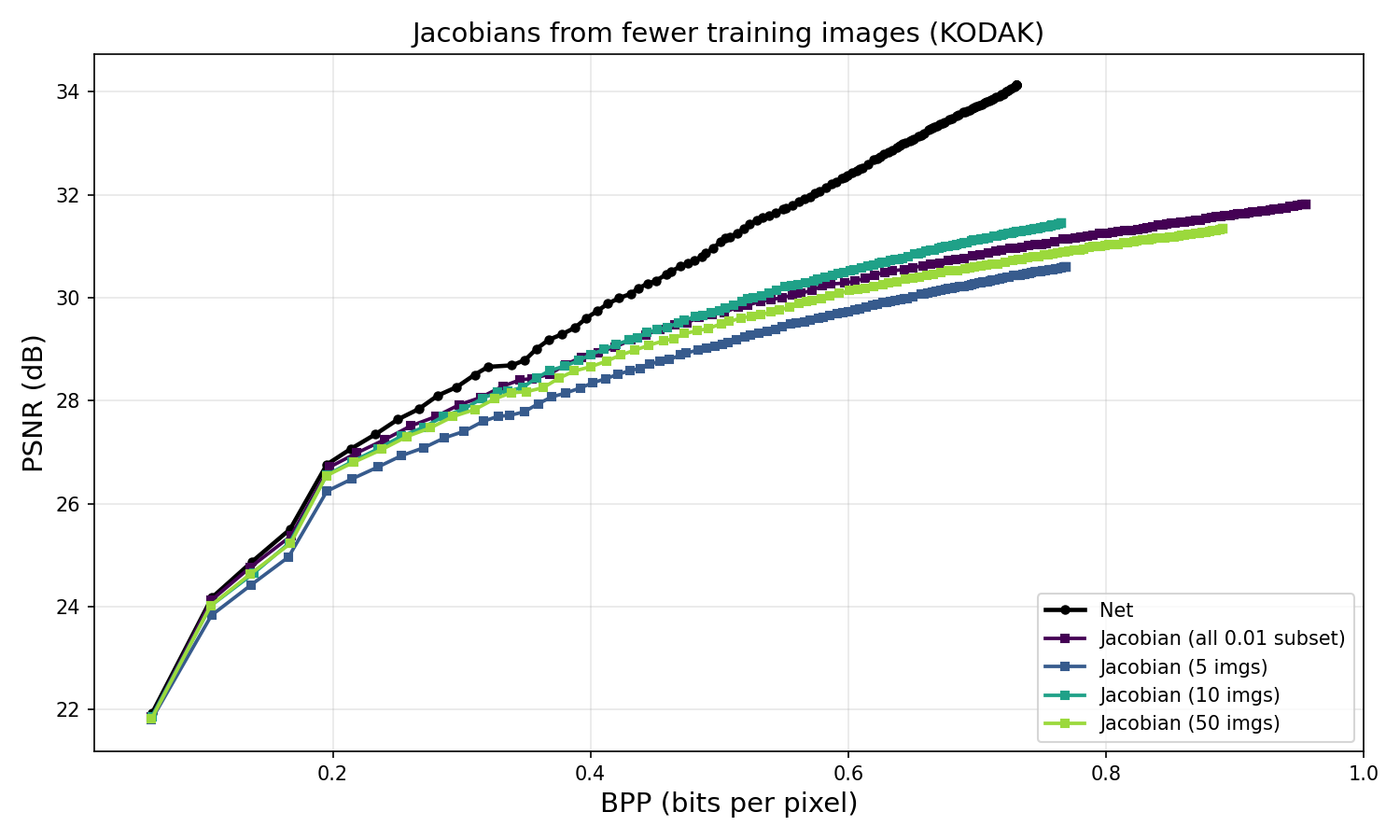}
\caption{\small{\textbf{Averaged Jacobians ablation.} PSNR vs. BPP on Kodak for the averaged Jacobians computed from 5, 10, 50 images and the full subset, compared with the original network.}}
\label{fig:jacobian_ablation}
\end{figure}

\begin{figure*}[!ht]
    \centering
    \begin{subfigure}{\linewidth}
        \centering
        \includegraphics[width=0.85\linewidth]{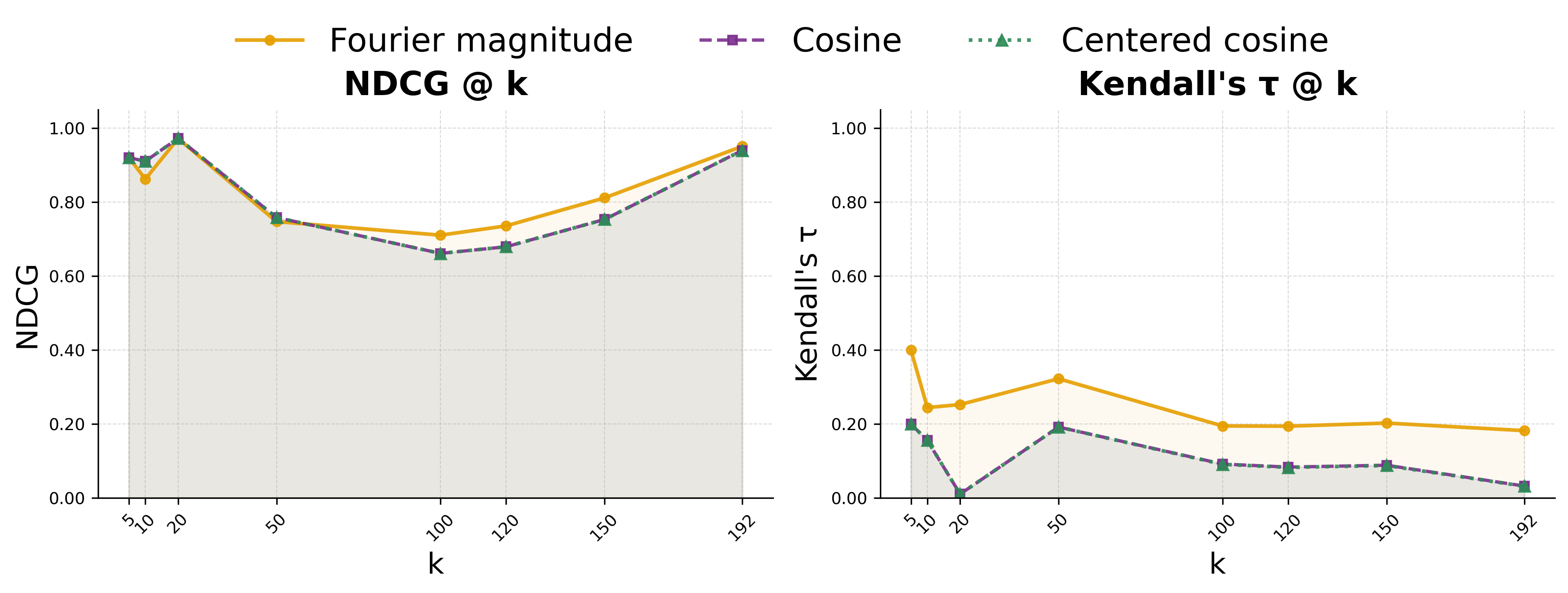}
        \caption{$\lambda = 0.05$}
        \label{fig:measures_l005}
    \end{subfigure}

    \vspace{0.6em}

    \begin{subfigure}{\linewidth}
        \centering
        \includegraphics[width=0.85\linewidth]{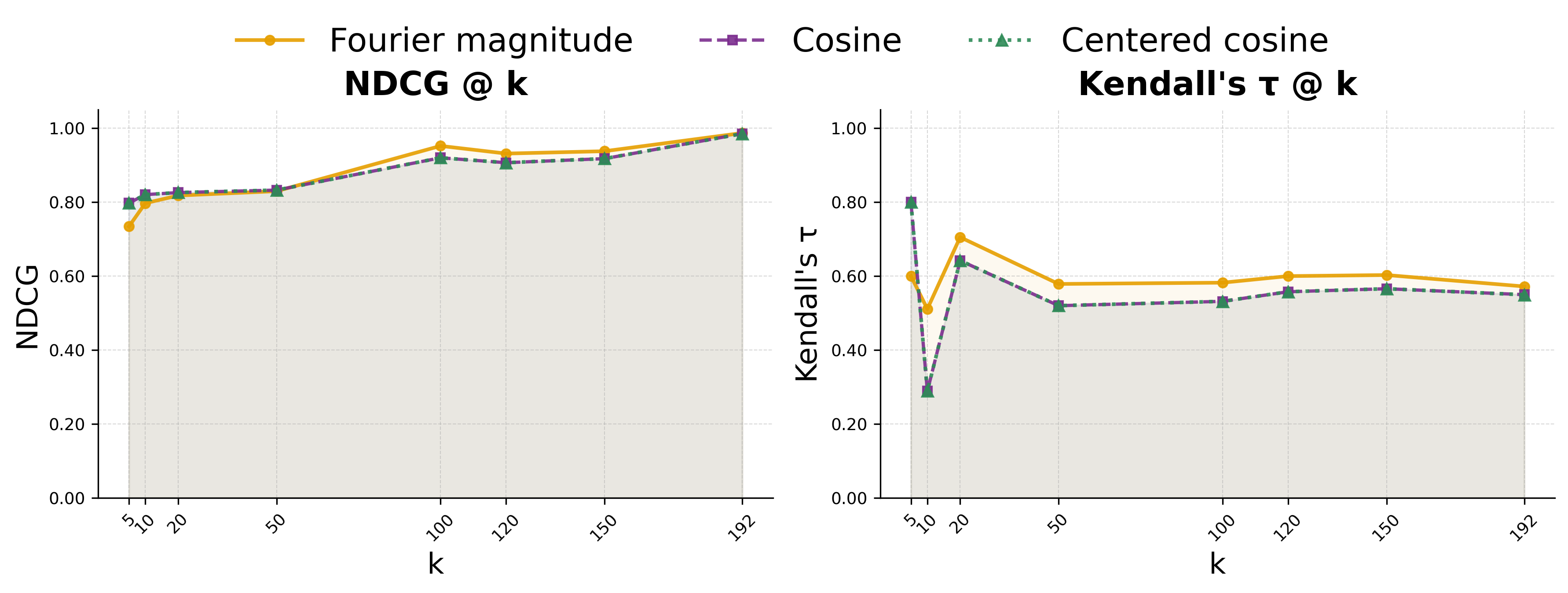}
        \caption{$\lambda = 0.8$}
        \label{fig:measures_l08}
    \end{subfigure}

    \vspace{0.6em}

    \begin{subfigure}{\linewidth}
        \centering
        \includegraphics[width=0.85\linewidth]{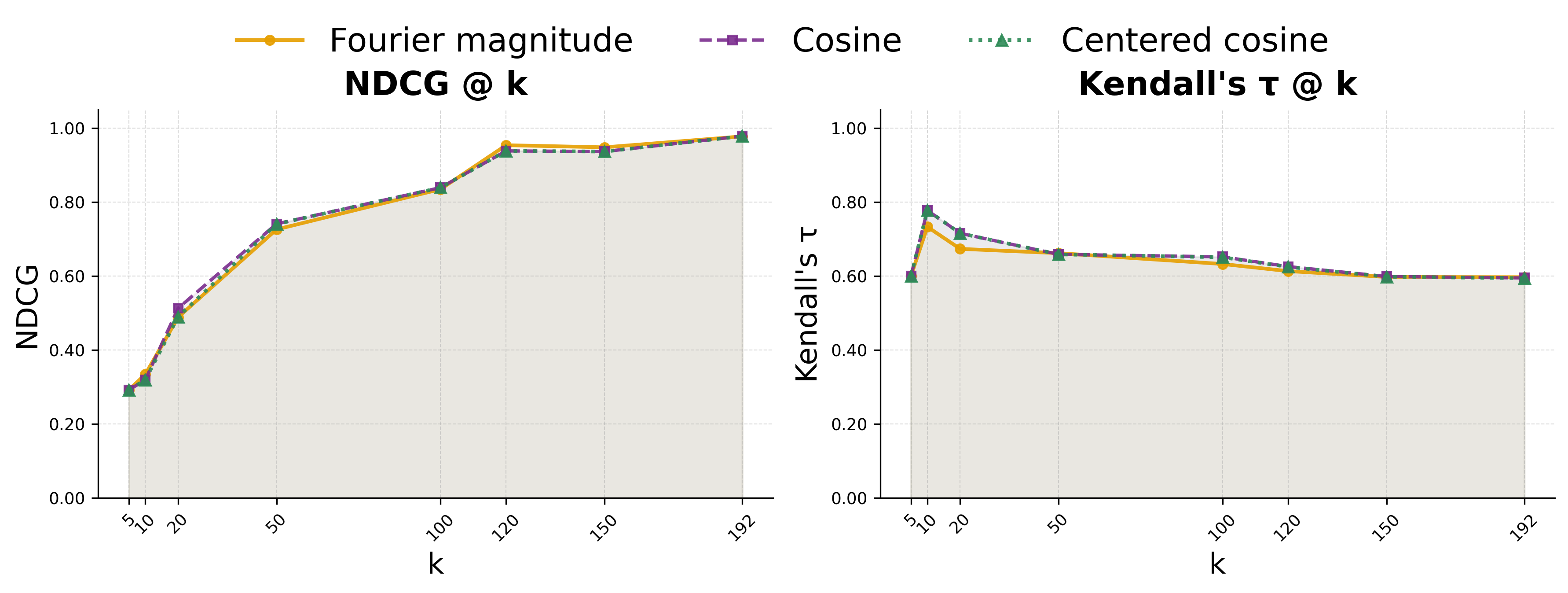}
        \caption{$\lambda = 2.0$}
        \label{fig:measures_l20}
    \end{subfigure}

    \caption{\small{\textbf{Similarity metric comparison}. Agreement between the energy-based feature ordering and the orderings obtained with
    different filter similarity measures: Fourier magnitude correlation, spatial cosine similarity,and centred cosine similarity (spatial normalized correlation). Left: NDCG@$k$; right: Kendall's
    $\tau$@$k$.}}
    \label{fig:similarity_measures}
\end{figure*}

\textbf{G. Ablation on correlation analysis:} The correlation analysis is performed using patches of size $16 \times 16$, corresponding to the receptive field of the encoder module and producing spatially centered Jacobian filters. Consequently, we would expect the phase to have a limited impact on the analysis, as the filters are spatially centered. To investigate this, Fig.~\ref{fig:similarity_measures} compares the ordering obtained using the Fourier magnitude, as presented in the paper, with two additional spatial-domain metrics: cosine similarity and centered cosine similarity. The results exhibit similar trends across all three models, suggesting that, in this setting, the phase has a limited impact on the resulting ordering.

\textbf{H. Ablation on averaging Jacobians:} We use the average Jacobian to test the hypothesis of universality, not as an optimal way to combine Jacobians. If the filtered signal remains similar across patches and only its position changes predictably, averaging the Jacobians should not cause a significant loss in performance. This simple approach provides insights into the universality of the Jacobians. Fig.~\ref{fig:jacobian_ablation} shows the impact of the number of images used to compute the averaged Jacobians.

\begin{figure}[!hb]
\centerline{\includegraphics[width=0.9\linewidth]{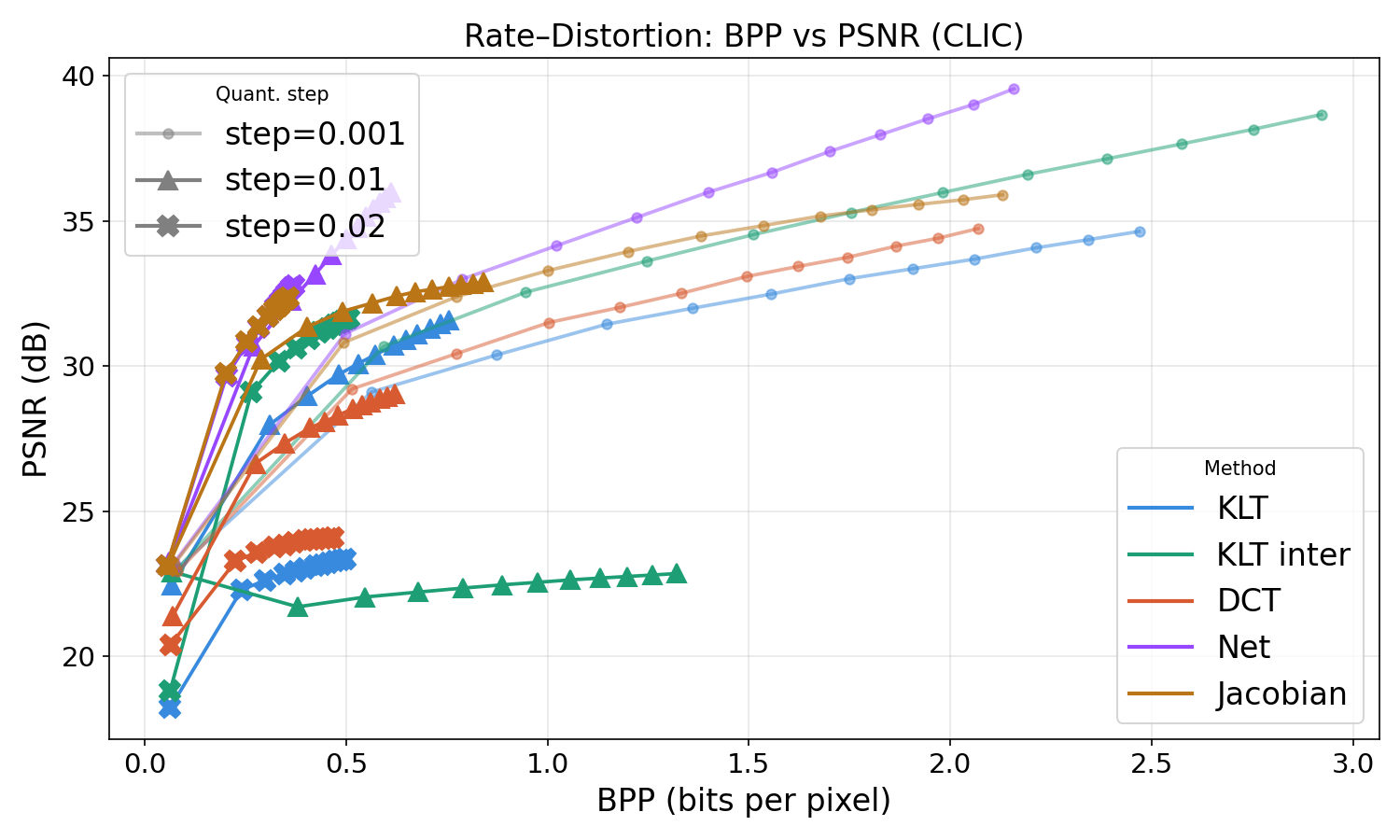}}
\caption{\small{\textbf{Comparison of different quantization steps and increasing the number of features for CLIC dataset.} For each method, we progressively increase the number of retained features according to their importance. We evaluate the original model, the Jacobian-based approach, and the transform-based methods DCT and KLT, considering separate and joint channel analysis (inter-channel) for KLT. Our Jacobian filtering method outperforms the transform-based and achieves PSNR values comparable to the original model, particularly at low BPP rates.}}
\label{fig:clic_comp}
\end{figure}

\textbf{I. Experiment on CLIC dataset:} We include the same evaluation from Fig.~\ref{fig:quantization_features} on CLIC dataset in Fig.~\ref{fig:clic_comp}. Which presents similar trends.

\begin{figure}[!hb]
\centerline{\includegraphics[width=0.9\linewidth]{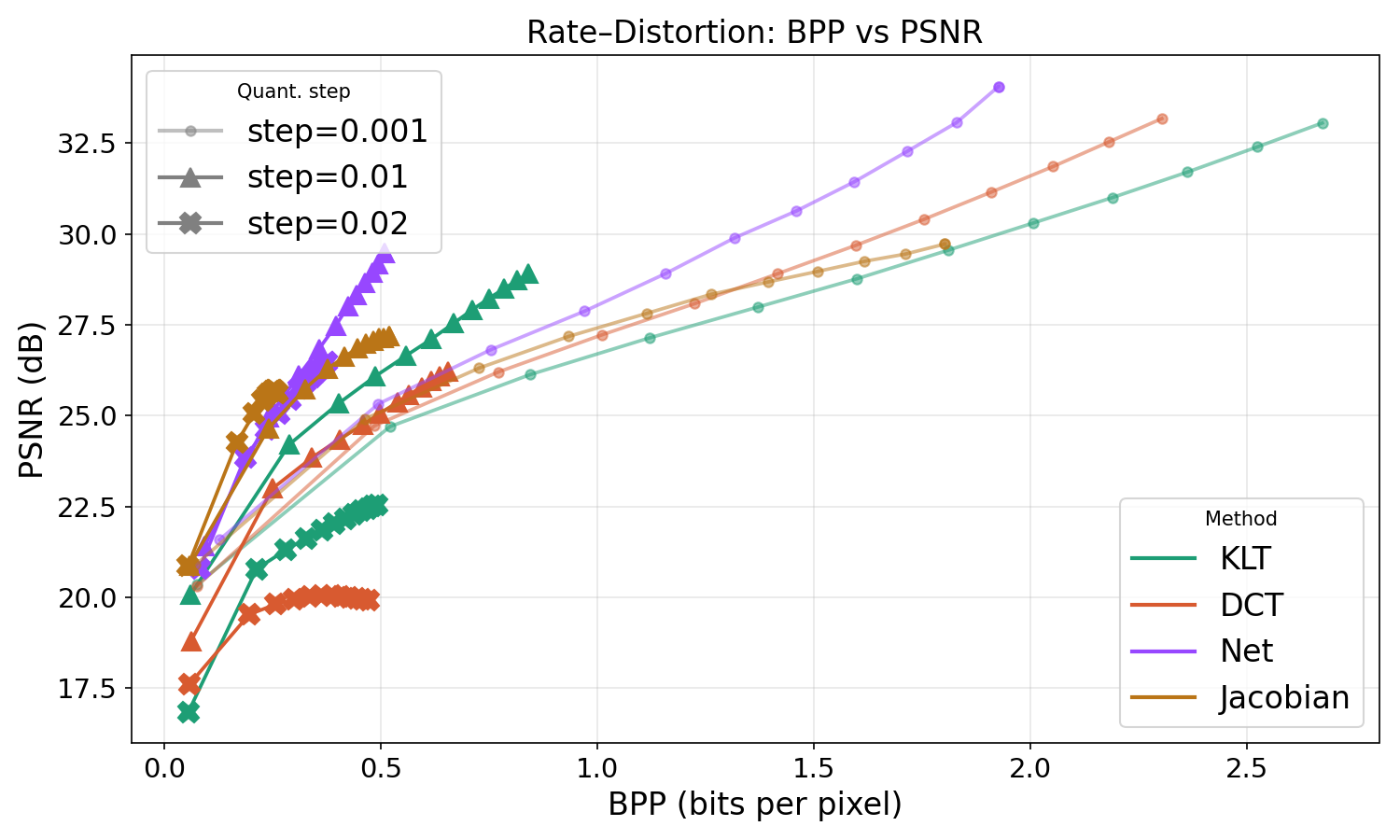}}
\caption{\small{\textbf{Single channel compression.} We compare three quantization steps, increasing the number of features according to importance, when compressing \textit{grayscale} images. We evaluate the original model, the Jacobian-based approach, and the transform-based methods DCT and KLT. Jacobian and Net models achieve better rate-distortion performance even in the absence of inter-channel correlations.}}
\label{fig:quantization_features_1ch}
\end{figure}

\textbf{J. Ablation on channel impact:} Since the inter-channel KLT achieves a better rate-distortion trade-off than the other transform-based methods at higher quantization steps ($q = 0.2$), we investigate whether the performance of the neural network and Jacobian models originates from their ability to use inter-channel correlations. To this end,  we train a model to compress grayscale images to remove the contribution of learned channel relationships. The results in Figure~\ref{fig:quantization_features_1ch} (same setup as in Section~\ref{sec:comp_3ch}) confirm that the neural network and Jacobians maintain a better rate-distortion trade-off even in a single-channel setting, suggesting that gains in performance cannot be attributed only to inter-channel correlations.

\begin{figure}[!hb]
\centerline{\includegraphics[width=0.9\linewidth]{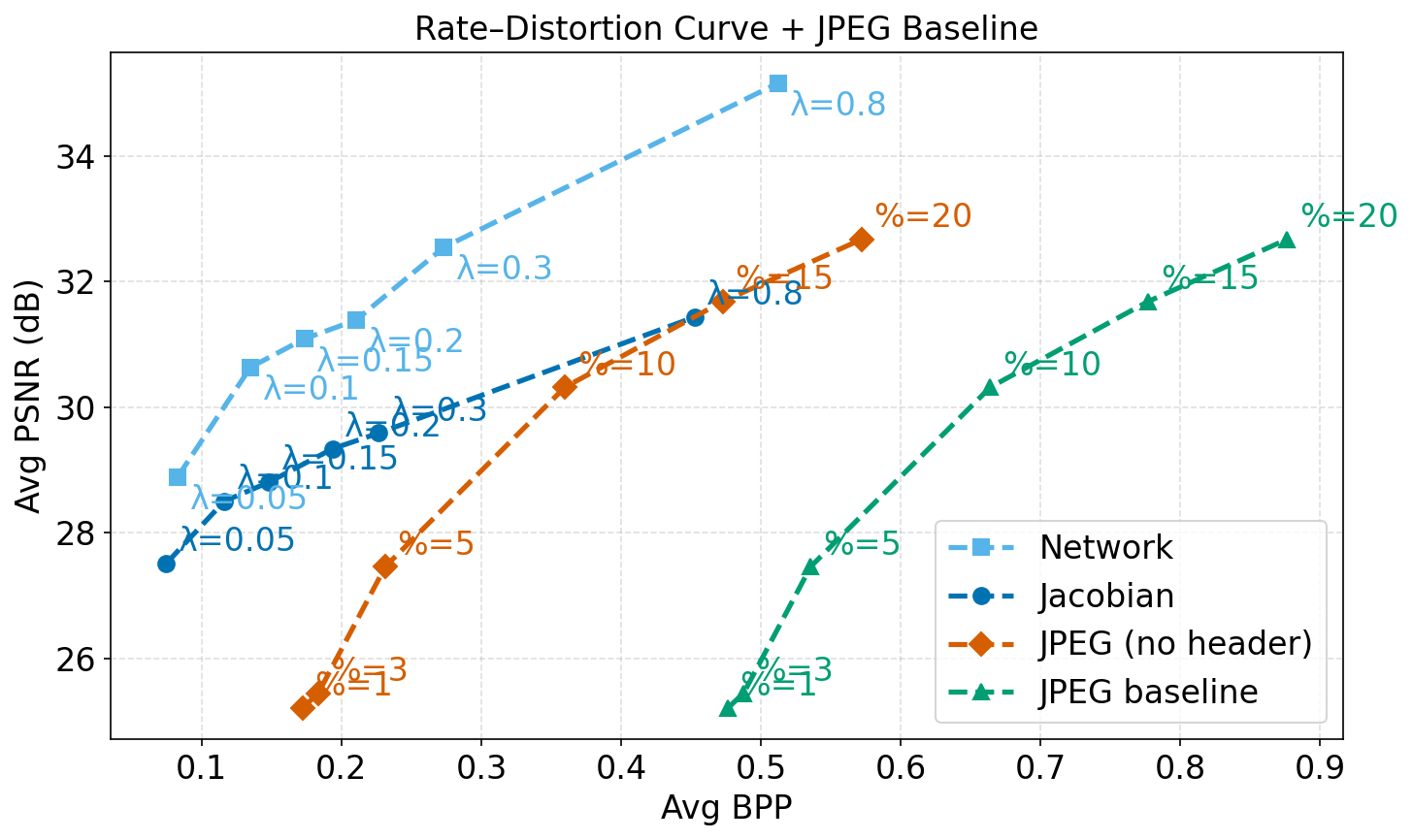}}
\caption{\small{\textbf{Comparison with JPEG compression.} We evaluate PSNR vs. BPP for six rate–distortion models ($\lambda$) and their Jacobian variants, and compare them with JPEG at different quality levels, with and without headers. The Jacobian approach outperforms JPEG in most settings, especially at low bit rates, but needs more sophisticated combination strategies to match the original model's performance.}}
\label{fig:comp_jpeg}
\end{figure}

\textbf{K. Additional Rate-Distortion experiments:} To evaluate our combination approach as a low-complexity image compression model, we compare models trained with different distortion parameters, $\lambda \in \{0.05, 0.1, 0.15, 0.2, 0.3, 0.8\}$. For each model, we select the corresponding $k^*$ and compute Jacobians for each feature representation. For a more realistic scenario, we use the learned entropy model associated with each trained model for both the network-based and Jacobian-based approaches. We use JPEG compression from Pillow library. We compare the BPP between Network, Jacobian, and JPEG compression, with and without header overhead.

\textit{PSNR vs. BPP:} Figure~\ref{fig:comp_jpeg} shows that, despite using a naive combination of Jacobians (Section~\ref{sec:combination}), our approach outperforms both JPEG variants, particularly in lower-rate. However, replacing all sufficient network features with Jacobians, while promising for low-complexity coding, can degrade performance, as shown in Figures~\ref{fig:quantization_features} and~\ref{fig:quantization_features_1ch}. This suggests that more sophisticated or hybrid strategies are needed.

\end{document}